\documentclass{mn2e}
\usepackage{graphicx}
\usepackage{amsbsy}
\usepackage{amsmath}
\usepackage{epsf}

\newcommand{\apj}{ApJ}

\newcommand{\mnras}{MNRAS}

\newcommand{\aap}{A\&A}
\newcommand{\qjras}{QJRAS}
\newcommand{\nat}{Nature}
\newcommand{\rvmp}{RvMP}

\newcommand\bb[1] {\mbox{\boldmath{$#1$}}}
\newcommand\del{\bb{\nabla}}
\newcommand{\Alf}{Alfv\'{e}n}

\newcommand{\beq}{\begin{equation}}
\newcommand{\eeq}{\end{equation}}

\DeclareMathAlphabet{\mathsfsl}{OT1}{cmss}{bx}{sl}
\SetMathAlphabet{\mathsfsl}{bold}{OT1}{cmss}{bx}{sl}

\usepackage{epstopdf}
\begin{document}

%\newcommand{\etal}{{\it et al.}}
%\newcommand{\beq}{\begin{equation}}
%\newcommand{\eeq}{\end{equation}}
%\newcommand{\mg}{M_{\rm g}}
%\newcommand{\ms}{M_*}
%\newcommand{\msol}{{\rm M_{\odot}}}
%
%\begin{document}

%===========================================================================
\title[B-$\rm \rho$ relation revisited]
{Magnetic Field -- Gas Density Relation and Observational Implications Revisited}
%===========================================================================

\author[Tritsis et al.]
  {A.~Tritsis$^{1}$, G.~V.~Panopoulou$^{1}$,   T.~Ch.~Mouschovias$^3$, K.~Tassis$^{1,2}$, V. Pavlidou$^{1,2}$ \\
    $^1$Department of Physics, University of Crete, PO Box 2208, 71003 Heraklion, Greece\\
    $^2$IESL, Foundation for Research and Technology-Hellas, PO Box 1527, 71110 Heraklion, Crete, Greece \\
    $^3$Departments of Physics and Astronomy, University of Illinois at Urbana-Champaign, 1002 W. Green Street, Urbana, IL 61801}

\maketitle 

%\label{firstpage}
\begin{abstract}
We revisit the relation between magnetic-field strength ($B$) and gas density ($\rho$) for contracting  interstellar clouds and fragments (or, cores), which is central in observationally determining the dynamical importance of magnetic fields in cloud evolution and star formation. Recently, it has been claimed that a relation $B \propto \rho^{2/3} $ is statistically preferred over $B \propto \rho^{1/2}$ in molecular clouds, when magnetic field detections and nondetections from Zeeman observations are combined. This finding has unique observational implications on cloud and core geometry: The relation $B \propto \rho^{2/3} $ can only be realized under spherical contraction. However, no indication of spherical geometry can be found for the objects used in the original statistical analysis of the $B-\rho$ relation. We trace the origin of the inconsistency to simplifying assumptions in the statistical model used to arrive at the $B\propto \rho^{2/3}$ conclusion and to an underestimate of observational uncertainties in the determination of cloud and core densities. We show that, when these restrictive assumptions are relaxed, $B \propto \rho^{1/2}$ is the preferred relation for the (self-gravitating) molecular-cloud data, as theoretically predicted four decades ago.
\end{abstract}

\begin{keywords}
diffusion -- ISM: magnetic fields -- ISM: clouds -- MHD -- stars: formation -- methods: statistical
\end{keywords}

\section{Introduction}\label{intro}

\begin{figure*}
\includegraphics[width=1.95\columnwidth, clip]{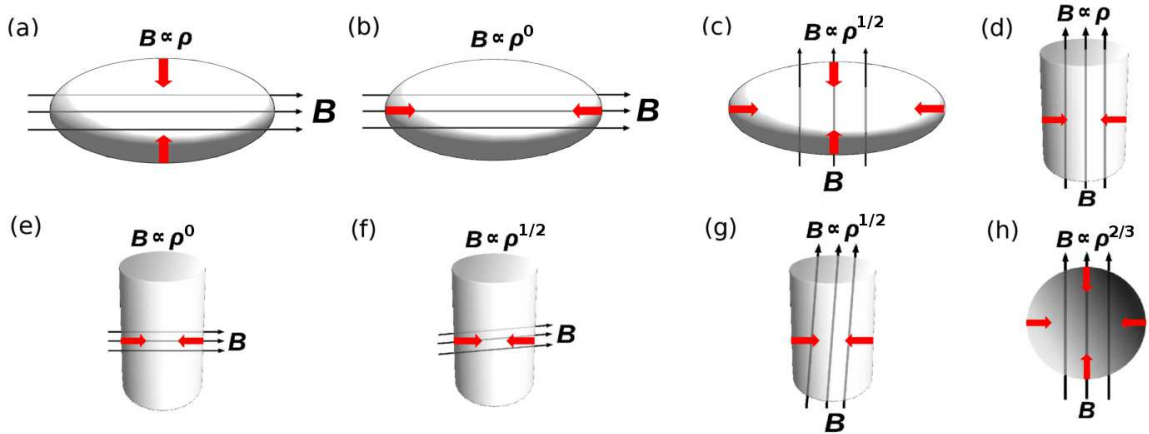}
\caption{Different geometries of contracting clouds and magnetic fields examined in \S \ref{geosection}. Black arrows represent the direction of the magnetic field and bold red arrows the direction of contraction. The $B \propto \rho^{2/3} $ relation is uniquely associated with spherical contraction and, therefore, has unique observational implications.
\label{geometries}}
\end{figure*}
Whether interstellar magnetic fields play a role in the formation of clouds and stars or are affected by cloud and/or star formation are old questions that predate the discovery of molecular clouds. Mestel \& Spitzer (1956) were concerned that the estimated very high electrical conductivity of the interstellar gas implies that the magnetic flux is frozen in the matter and, in the then prevailing picture of star formation (Hoyle's spherical collapse and hierarchical fragmentation), magnetic fields would prevent fragmentation and star formation. For this reason, they suggested that ambipolar diffusion (the motion of electrons and ions together with magnetic flux relative to the neutrals) would set in at some stage and allow a cloud to reduce its magnetic flux and thereby fragment and collapse to form stars.

The first testable prediction of magnetic-field strengths in clouds destined to form stars was given by Mestel (1965). A cloud collapsing spherically and isotropically while conserving its mass ($M$) and magnetic flux ($\Phi_B$) implies a relation between the magnetic-field strength and gas density: $M \propto \rho R^3 = const$, $\Phi_B \propto B R^2 = const'$. Hence, eliminating the cloud radius $R$ in favor of the density $\rho$ from the two conservation laws yields a scaling between the field strength $B$ and the density: $B \propto \rho^{2/3}$.

Verschuur (1969) summarized the results of Zeeman measurements of the field strength in nine H\small{I} clouds on a log($B$) -- log($n$) plot, where $n$ is the number density (particles/${\rm cm^3}$) and concluded, as did other workers, that the measurements were in agreement with that theoretical prediction -- without, however, performing an actual fit to those data. 

Mouschovias (1976a, b) provided the first self-consistent formulation and solution of the problem of the equilibrium of self-gravitating, isothermal, magnetic clouds embedded in a hot and tenuous intercloud medium. He also considered the implication of the contraction of such clouds (or fragments) on the $B$ -- $\rho$ relation. In the deep interior of {\it each} cloud, he found that the ratio of magnetic and gas pressures, $\alpha \equiv B^2/8\pi P$, tends to remain constant during contraction. In fact, it tends to retain a value near unity. For isothermal contraction, $P = \rho_{\rm n}C^{2}$, where $C = (k_{\rm B}T/{\bar{m}})^{1/2}$ is the isothermal speed of sound at temperature $T$ and mean mass per particle $\bar{m}$; the quantity $k_{\rm B}$ is Boltzmann's constant. Hence, Mouschovias' result $\alpha \approx 1$, specialized to isothermal contraction, yields $B \propto \rho^{1/2}$. For an equation of state $P \propto \rho^{\gamma}$, the relation between $B$ and $\rho$ becomes $B \propto \rho^{\gamma /2}$ (see review by Mouschovias 1991b). An analysis by Crutcher (1999) of a larger sample of clouds than that used by Verschuur, with measured magnetic-field strengths and number densities (greater than 100 $\rm cm^{-3}$), found a best-fit exponent of $0.47$ for the $B-\rho$ scaling, in agreement with the theoretical prediction. Detailed numerical simulations by Fiedler \& Mouschovias (1992, 1993) predicted a slope $\kappa = 0.47$ for contracting cores formed by gravitationally-driven ambipolar diffusion and evolving from initially magnetically subcritical to supercritical states. More recently, Li et al. (2015) inferred $\kappa = 0.41$ from observations in the massive star forming region NGC 6334. For lower densities, although the theoretical prediction for an evolving cloud was $\kappa \approx 0$ (because self-gravity is not strong enough to compress the cloud perpendicular to the field lines), the observational picture has generally been less clear, with various scaling exponents derived empirically (see Vall\'{e}e 1997 and references therein; Marchwinski 2012) for sets of clouds observed at the same time. 

Crutcher et al. (2010, hereinafter referred to as CWHFT10) revisited the scaling between $B$ and $\rho$ for a yet larger sample of both low-density (primarily H\small{I}) and high-density (primarily molecular) clouds. They used a Bayesian statistical analysis that allowed them to treat nondetections and varying angles between the magnetic field and the line of sight, and they optimized a family of models consisting of a uniform distribution of magnetic-field values between some minimum and maximum, with the maximum having two distinct branches in its behavior (on a log$B$ -- log{$\rho$} plot): a flat part at low densities ($B$ independent of $\rho$), and a power-law scaling at higher densities ($B\propto\rho^\kappa$), with the exponent $\kappa$, the break density $\rho_0$, and the width of the magnetic-field strength distribution being the free parameters of their model. Their conclusion was that the data prefer $\kappa \approx 2/3$ and reject $\kappa \approx 1/2$. They took this result to be an indication of ``isotropic contraction of gas too weakly magnetized for the magnetic field to affect the morphology of the collapse.''

In this work, we examine more closely the observational implications of different geometries of contraction on the $B$ -- $\rho$ relation. The distribution of forces in a cloud determines its evolution, including its geometric shape and the associated $B$ -- $\rho$ relation. Although a given (or observed) $B$ -- $\rho$ relation does not necessarily imply a unique geometric shape of a cloud, it is nevertheless the case that a given (or observed) $B$ -- $\rho$ relation can only be found in a very restricted set of geometric shapes, which in turn restrict the kind of motions capable of producing those shapes and the $B$ -- $\rho$ relation. Here, we test whether the observed shapes of the objects (clouds and cores) on which the latest $B$ -- $\rho$ relation study (that of CWHFT10) has been based are consistent with the underlying geometries in which the claimed scaling ($B \propto \rho^{2/3}$) could develop. 

In \S \ref{geosection} we summarize the $B$ -- $\rho$ relations implied by different cloud geometries that could be established by the evolution of molecular clouds with frozen-in magnetic fields (no significant ambipolar diffusion). Density maps of clouds and cores used in CWHFT10 are examined in \S \ref{obshapes}, testing for consistency between geometry and the exponent $\kappa$. The value $\kappa = 2/3$ claimed by CWHFT10 cannot be reconciled with the observed cloud shapes. The source of the discrepancy lies in various assumptions of the CWHFT10 analysis, as we show in \S \ref{tracing}. Relaxing the problematic assumptions, we reconcile the observed shapes and the $B$ -- $\rho$ relation in \S \ref{reconciling}, and we show that the value $\kappa = 1/2$ is preferred by the data over the value $\kappa = 2/3$. We summarize the conclusions in \S \ref{sum}.

\section{Cloud Geometry and the $B$ -- $\rho$ Relation}\label{shapes}

In this section we address the connection between the slope of the $B$ -- $\rho$ relation and the cloud geometry. First, in \S \ref{geosection}, we investigate theoretically the $B$ -- $\rho$ relation implied by different geometries of clouds and magnetic fields. Then, in \S \ref{obshapes}, we examine the shapes of objects in the CWHFT10 sample and whether they are consistent with the claimed slope $\kappa = 2/3$.

\subsection{$B$ -- $\rho$ Relations Implied by Different Geometries}\label{geosection}

\subsubsection{Disklike or Slab Cloud with \bb{B} in the Plane of the Disk}

We first consider an oblate (disklike) cloud of half-thickness $Z_0$ and arbitrarily large radius $R_0$, uniform density $\rho_0$, threaded by a uniform magnetic field $B_0$ in the plane of the disk (see Fig. 1a). A slab-shaped cloud is a special case of this. Contraction perpendicular to the plane of the disk to a new half-thickness $Z$ increases the density and the magnetic field by the same factor, $Z_{0}/Z$; hence, $B \propto \rho$. One should note that, for this kind of contraction, the gravitational force per unit mass perpendicular to the plane of the disk on a fluid element initially at $z$ depends only on the column density $\rho z$, which does not change upon contraction. However, the magnetic-pressure force per unit mass on that same fluid element, $|-\del{B^2/8\pi}|/\rho$, increases upon contraction as $(Z_{0}/Z)^2$. Consequently, even if such contraction sets in for some reason, magnetic forces will eventually stop it.

We now consider the same oblate cloud, threaded by the same magnetic field, but now the cloud is allowed to contract only along the field lines (see Fig. 1b); i.e., the half-thickness (or polar radius) $Z_0$ does not change, but the local extent of the cloud along field lines ($\propto R_0 \cos{\theta}$, where $\theta$ is the angle between the field lines and a line from the cloud's centre to the point of interest on the rim of the cloud) decreases such that the density increases uniformly in the cloud model. The ultimately resulting shape is in general one of a prolate, triaxial object, a ``filament" perpendicular to the field lines. The motions that created this filamentary cloud do not by themselves change the strength of the magnetic field; hence, $B \propto \rho^0$, i.e., $B$ is independent of $\rho$. However, the increased density in the filamentary structure implies a stronger gravitational field {\it along} the filament, toward the centre of the original oblate cloud. {\it Will this filament fragment along its length to form at least one more-or-less spherical core?}

If this filament contracts as a whole along its length, an argument similar to the one in Mouschovias (1976b) shows that the magnetic-tension force near the ends of the filament increases more rapidly than the gravitational force, so such contraction (perpendicular to the field lines) can more easily take place in the central part of the filament. If a fragment (or core) is to separate out and contract gravitationally in this region, its mass-to-flux ratio must exceed the critical value for collapse, 
\begin{equation}
\label{m2f}
\mu_{\rm cr} \equiv \left(\frac{M}{\Phi_{\rm B}}\right)_{\rm cr} = \left(\frac{\sigma}{B}\right)_{\rm cr} \approx \frac{1}{\sqrt{63 G}} \, 
\end{equation}
\noindent (Mouschovias \& Spitzer 1976). The quantity $\sigma$ is the column density along field lines (in g ${\rm {cm}^{-2}}$), and $B$ is the magnetic field strength. (The constant on the right-hand side of eq. [1] has a slight dependence on the geometry of the cloud.) If the original size of the cloud, both parallel and perpendicular to the field lines, is very large, then the resulting filament will also be very long, and it is possible for several fragments to separate out along its length, provided that criterion (1) is satisfied for each. The thermal critical mass per unit length of a filament, $2C^2/G$ (Ostriker 1964), is not a relevant quantity for the fragmentation of a filament threaded by a magnetic field perpendicular to its long dimension. For each fragment, $B \propto \rho^{1/2}$ for as long as the magnetic field remains frozen in the matter. Detailed numerical simulations (Fiedler \& Mouschovias 1992, 1993) showed that ambipolar diffusion sets in in the interiors of initially subcritical molecular clouds and leads to an increase in the mass-to-flux ratio toward its critical value for collapse (see eq. [1]). Prior to establishment of critical conditions, the magnetic-field strength increases by at most 30\% while the density can increase by a very large factor; hence, $\kappa \approx 0$. After the mass-to-flux ratio reaches its critical value, contraction accelerates and proceeds with balance of forces along field lines and as rapidly as magnetic forces allow perpendicular to the field lines. These are the sufficient conditions for establishment of the relation $B \propto \rho^{\kappa}$ with $\kappa = 1/2$. (Actually, the numerical simulations show that $\kappa = 0.47$, meaning that flux-freezing is not perfect; some ambipolar diffusion takes place even during the dynamical stage of contraction -- see Fiedler \& Mouschovias 1993, Fig. 9c.)

\subsubsection{Disklike or Slab Cloud with \bb{B} Perpendicular to the Plane of the Disk}

We now consider the same disklike cloud, threaded by the same magnetic field as above, but now the field is perpendicular to the plane of the disk (see Fig. 1c), with field lines ``fanning out'' outside the cloud, acquiring a characteristic hour-glass shape (not shown in Fig. 1c because it is not essential for the present purposes). Such oblate clouds are unavoidable if they form out of lower-density interstellar gas in which the magnetic field has an ordered component and, locally, the magnetic force is nonnegligible relative to the gravitational force. The $B \propto \rho^{1/2}$ relation implied by the gravitational, isothermal contraction of such a cloud (both along and perpendicular to the field lines) as well as its physical origin were described above in the {\it Introduction} and at the end of the preceding subsection and they need not be repeated here. Fragments can separate out in the interior of the cloud and contract dynamically if their mass-to-flux ratio exceeds the critical value for collapse given by equation (1). A magnetically subcritical cloud can reach critical conditions because of gravitationally-driven ambipolar diffusion, whose modern understanding is that it redistributes mass in the central flux tubes of molecular clouds, where the degree of ionization is relatively small ($< 10^{-7}$), but it does not lead to flux loss by a cloud as a whole (Mouschovias 1979).

\subsubsection{Cylindrical or Filamentary Cloud with \bb{B} Along the Cylinder}

Cylindrical model clouds threaded by a magnetic field along the axis of symmetry of the cylinder (see Fig. 1d) were studied exhaustively by Mouschovias \& Morton (1991, 1992a, 1992b). During lateral contraction (perpendicular to the symmetry axis and the magnetic field), both the instantaneous density $\rho$ and magnetic-field strength $B$ of a fluid element at instantaneous cylindrical polar radius $r$ increase as $1/r^2$; hence, $B \propto \rho$. The gravitational force per unit mass at the position of the fluid element increases as $1/r$, while the magnetic force per unit mass increases as $1/r^3$. Consequently, the magnetic forces will stop such contraction at some stage. There is neither a critical mass per unit length nor a critical mass-to-flux ratio per unit length for such self-gravitating filaments to suffer collapse (see Mouschovias \& Morton 1991, discussion following eq. [38]). The evolution of such model clouds was followed numerically by Mouschovias \& Morton (1992a, b); the density quickly acquires a spatial profile approximated by $1/r^2$.

\subsubsection{Cylindrical or Filamentary Cloud with \bb{B} Perpendicular to the Cylinder}

The case in which the magnetic field is perpendicular to the axis of the cylinder (Fig. 1e) is similar to the one discussed in \S 2.1.1 above -- the object that started as an oblate (disklike) cloud threaded by a magnetic field in the plane of the disk and then contracted primarily along the field lines to acquire a triaxial, prolate shape perpendicular to the field lines.

\subsubsection{Cylindrical or Filamentary Cloud with \bb{B} at an Angle with respect to the Cylinder}

If the magnetic field is at an angle with respect to the axis of the cylindrical cloud, one might think that its component along the axis would ensure that there is no critical mass or mass-to-flux ratio for lateral collapse, while its component perpendicular to the axis would bring in the Mouschovias \& Spitzer (1976) critical mass-to-flux ratio for collapse and/or fragmentation of the cloud in both directions. Then the longitudinal component of $\bb{B}$ would tend to increase as $\rho$, while the lateral component would tend to increase as $\rho^{1/2}$. Unfortunately, the effect of the magnetic field on the evolution of a cloud cannot be deduced correctly by considering separately the effect of each of its components and then superimposing the two results. To visualize the behavior of a prolate cloud threaded by a magnetic field at an angle with respect to its length, we consider two cases: (a) $\bb{B}$ is almost perpendicular to the 
axis of the cylinder (see Fig. 1f), and (b) $\bb{B}$ is almost parallel to the axis (see Fig. 1g). 

In case (a), motions along the field lines can, in principle, take place and make the extent of the cloud along $\bb{B}$ as small as thermal-pressure and/or turbulent-pressure forces allow. If only this evolution took place, $B$ would be independent of $\rho$. However, for self-gravitating fragments (or cores) to form, destined to collapse and form stars, contraction {\it perpendicular} to the field lines has to take place as well. This brings in the critical mass-to-flux ratio given by equation (1), and the relation $B \propto \rho^{1/2}$ is established again.

In case (b), motions along field lines can transform the cylinder into an oblate (disklike) object or even break it up into several such oblate objects, with hardly affecting the magnetic-field strength. However, as we have already seen, for an oblate object to collapse its mass-to-flux ratio has to exceed the critical value. Evolution beyond that state results in the relation $B \propto \rho^{1/2}$. How the fragmentation of an initially critical magnetic flux tube affects the $B$ -- $\rho$ relation depends on the number of fragments that may form along the flux tube, and is discussed in detail in Mouschovias (1991c; see \S~2.4 and Fig. 1 therein). 

\subsubsection{Spherical Cloud}

The $B$ -- $\rho$ relation implied by spherical, isotropic contraction (Fig. 1h) was discussed in \S\ref{intro}. The relation $B \propto \rho^{2/3}$ is unique among $B$ -- $\rho$ relations in that only spherical contraction can cause it. If contraction along field lines is more rapid than perpendicular to the field lines, the exponent $\kappa$ in the relation $B \propto \rho^{\kappa}$ becomes less than 2/3. If the opposite is true, the exponent $\kappa$ becomes greater than 2/3. Thus the recent claim by CWHFT10 that Zeeman and density observations, taken in aggregate, yield $B \propto \rho^{2/3}$ has unique, observationally testable implications on the shapes of the observed objects (clouds or cores): they {\em must }be spherical.

The severe constraint on the geometry that can produce the $B \propto \rho^{2/3}$ relation does not get relaxed by a random component of \bb{B} dominating its ordered (or mean) component inside the observed object (cloud or core)\footnote{In any case, at core scales (smaller than about 0.1 pc), the magnetic field is both observed (e.g., Girart et al. 2006, Chapman et al. 2013) and theoretically expected (Mouschovias 1987a) to be dominated by an ordered component. The reason is that, because of magnetically-driven ambipolar diffusion, a random component of \bb{B} cannot be sustained on scales smaller than the Alfv\'en lengthscale, $\lambda_A = \pi v_A \tau_{\rm ni}$, where $v_A = B/(4 \pi \rho)^{1/2}$ is the {\Alf} speed in the neutrals, and $\tau_{\rm ni}$ is the slowing-down time of a neutral particle due to collisions with ions; for typical molecular clouds, $\lambda_A = 0.1$ pc (Mouschovias 1987a; 1991a). The timescale for straightening out field lines that are tangled on a scale {\it l} is proportional to $l^2$ and, for typical molecular cloud densities ($10^3 \, {\rm cm^{-3}}$), magnetic fields ($30 \, \mu{\rm G}$), and core sizes (0.1 pc), is much smaller than the free-fall time.  (e.g., see Mouschovias et al. 2011).} . While a completely random component of the magnetic field does not introduce a spatial anisotropy that would tend to destroy the spherical geometry and the $B \propto \rho^{2/3}$ relation, it does not enter the $B$ -- $\rho$ relation, and it also does not contribute to the observed line-of-sight $B$.
The $B$ -- $\rho$ relation refers to the mean $B$, and Zeeman observations measure that field's component along the line of sight.  

\subsection{Observational Determination of CWHFT10 Cloud Shapes} \label{obshapes}

The conclusion that the exponent $\kappa = 2/3$ requires spherical clouds and cores is difficult to reconcile with the CWHFT10 result at high densities. Although the shapes of molecular clouds and their fragments are an important part of the debate on the process that regulates cloud evolution and star formation, spherical clouds and cores are not a contender in either the theoretical or the observational arguments. 

\begin{figure}
\centering
\includegraphics[scale=0.35]{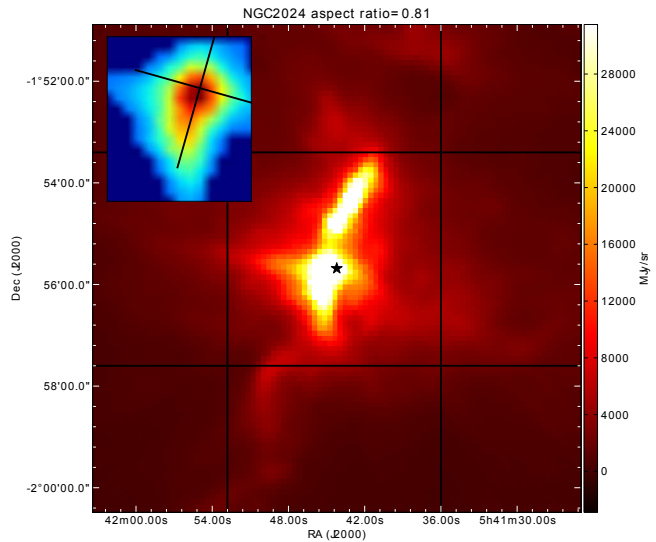}
\caption{Archival {\it Herschel}-SPIRE 250$\mu$m dust continuum emission map of the core NGC2024. The star represents the coordinates of the core as reported by Falgarone et al (2008). The inset image is the outcome of our data processing and the black lines represent the principal axes of the core.}
\label{NGC2024}
\end{figure}

\begin{figure*}
\centering
\includegraphics[scale=0.6]{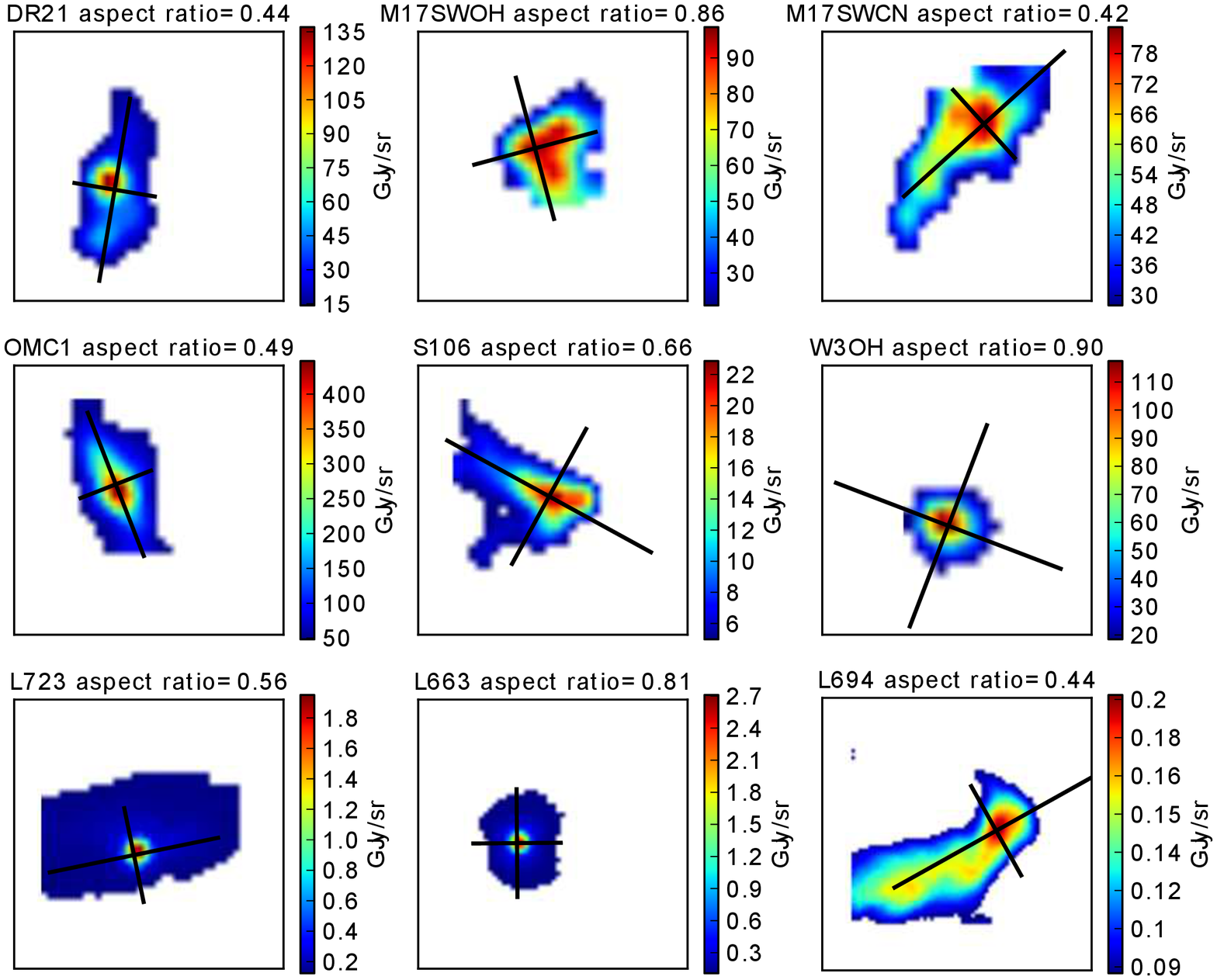}
\caption{250$\mu$m dust continuum emission maps from the {\it Herschel} Science Archive. Black lines represent the principal axes of the cores.}
\label{multipanel_maps_250mu}
\end{figure*}
\begin{figure*}
\centering
\includegraphics[scale=0.4]{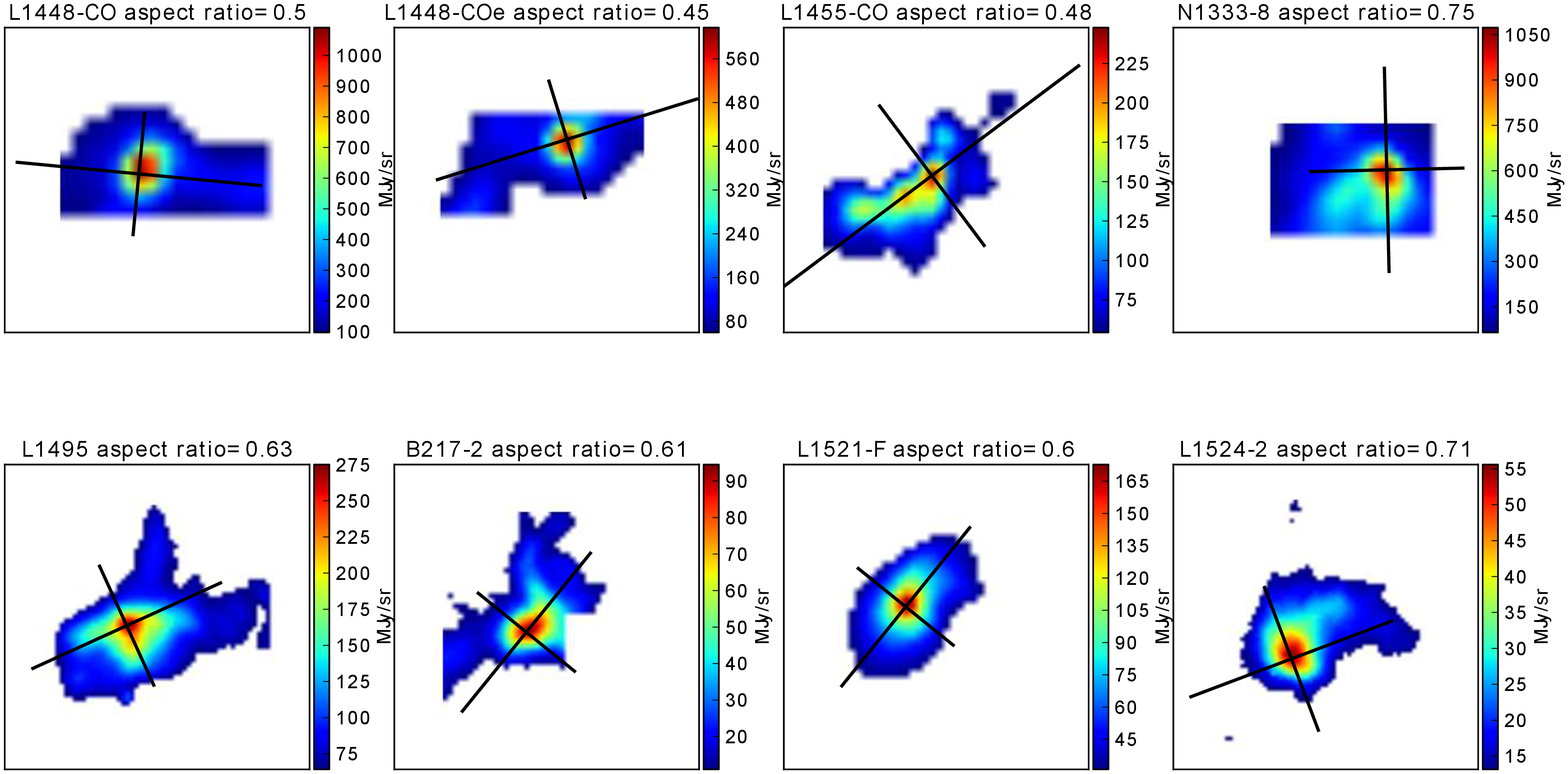}
\caption{500$\mu$m dust continuum emission maps from the {\it Herschel} Science Archive. Black lines represent the principal axes of the cores.}
\label{multipanel_maps_500mu}
\end{figure*}
\begin{figure*}
\centering
\includegraphics[scale=0.7]{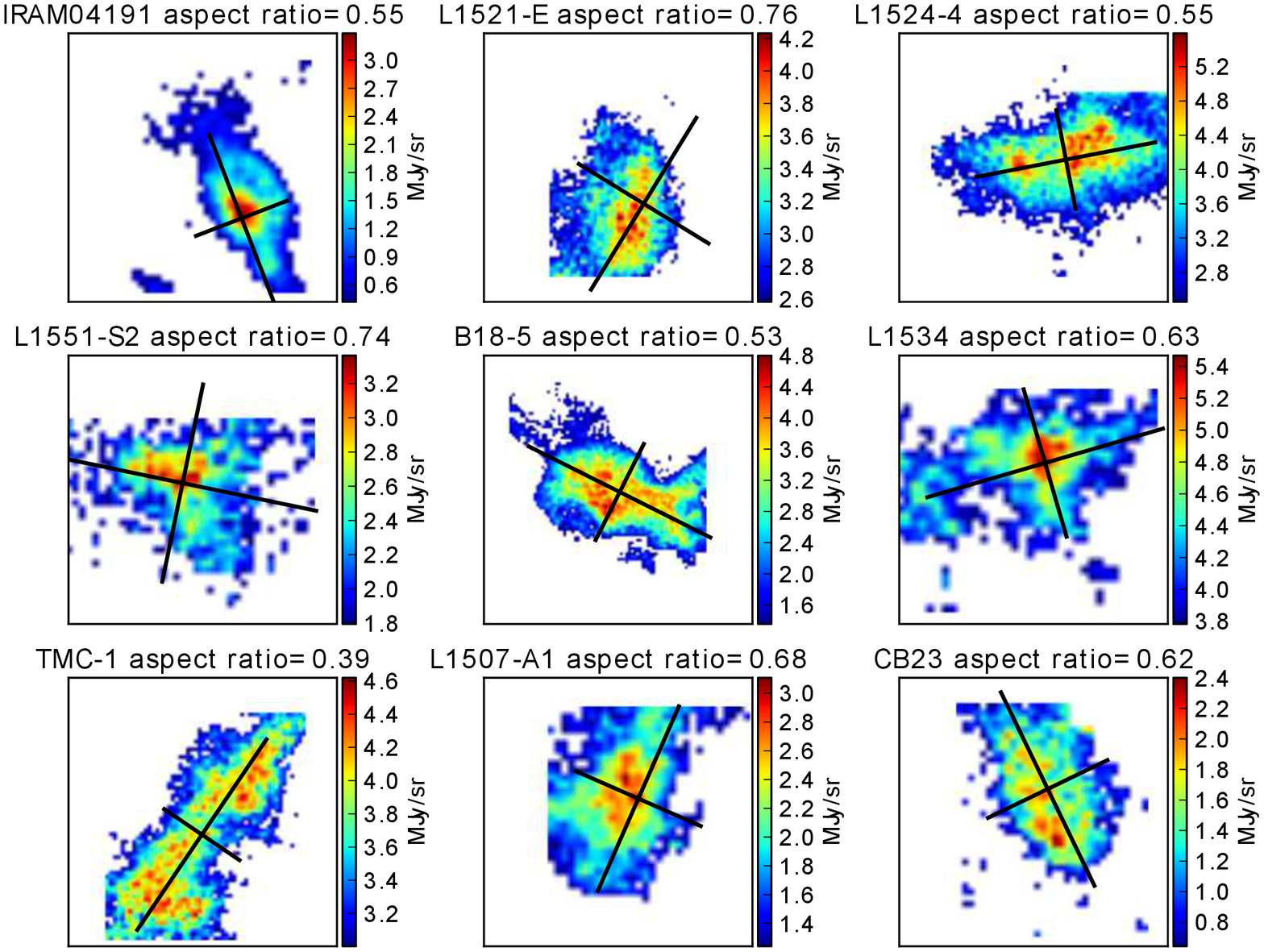}
\caption{$^{13}$CO line emission maps of the J=1-0 transition from the FCRAO survey. Black lines represent the principal axes of the cores.}
\label{multipanel_maps_CO}
\end{figure*}

According to theory, if ambipolar diffusion is the agent mainly responsible for core formation, then the cores are
expected to be flattened along the magnetic field because magnetic forces act only perpendicular to the field lines. Thus, the expected shapes are oblate (Mouschovias 1976b; Fiedler \& Mouschovias 1993), although not necessarily axisymmetric (Basu \& Ciolek 2004; Ciolek \& Basu 2006; Kudoh \& Basu 2011). If cores form as the result of converging turbulent flows (e.g., see review by Mac Low \& Klessen 2004), then they are expected to have random, triaxial shapes, with a slight preference for prolateness (Gammie et al. 2003; Li et al. 2004). Predominantly spherical cores and clouds are not expected by any formation mechanism. 

Observationally, the issue is how to distinguish between oblate and prolate intrinsic shapes from two-dimensional projections. Spherical clouds and cores are straightforward to spot even in projection: their two-dimensional projections are circles and the aspect ratio of the projected shapes is very sharply peaked at 1. 
Early work on core shapes, assuming axial symmetry, seemed to favour prolate cores (Myers et al. 1991; Ryden 1996). However, subsequent investigations which relaxed the axisymmetry assumption have consistently yielded triaxial, preferentially oblate core shapes (e.g., Jones, Basu \& Dubinski 2001; Jones \& Basu 2002; Goodwin, Ward-Thompson \& Whitworth 2002), independently of tracer or core sample. Tassis (2007) and Tassis et al. (2009) found strong indications for triaxial, preferentially oblate cores. Spherical cores and clouds on the other hand are not common in nature.

\begin{figure}
\centering
\includegraphics[scale=0.4, clip]{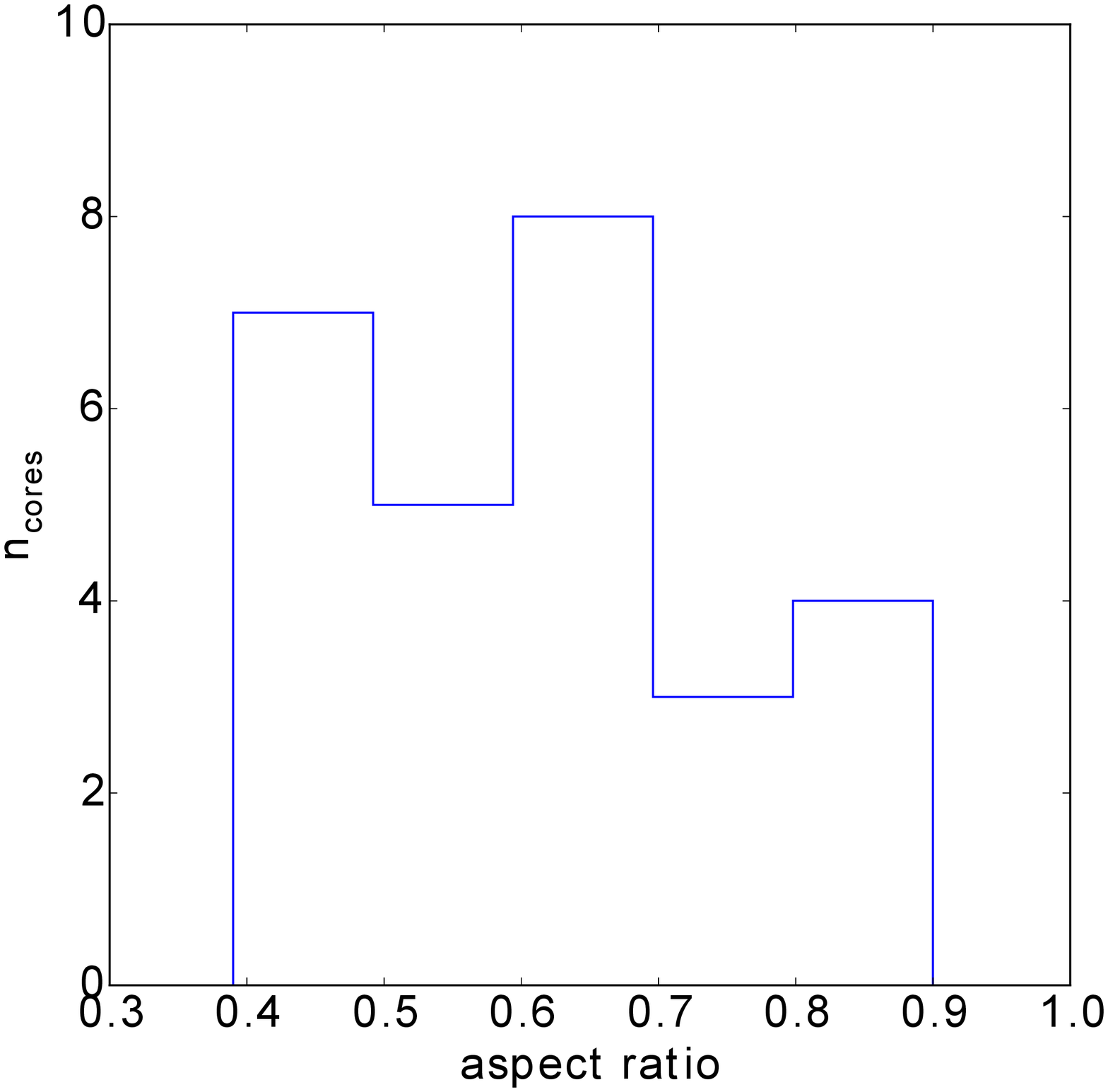}
\caption{Distribution of the aspect ratios of the cores for which we could find observational data. The distribution peaks away from unity, in contradiction to what is expected from spherical geometry.
}
\label{aspect_ratio_dist}
\end{figure}
However, it may still be possible that the molecular sample used in the CWHFT10 study does contain an unusually high fraction of spherical objects, which would be consistent with their finding of $B\propto \rho^{2/3}$. To test for this possibility, we have calculated aspect ratios of plane-of-the-sky maps for as many of the objects of CWHFT10 as we could find appropriate data.  

We used dust continuum emission images from the online data archive of the {\it Herschel} Space Telescope (Pilbratt et al. 2010) observed with SPIRE at 250 $\mu$m and 500 $\mu$m to create maps of 18 of these objects, and $\rm ^{13}CO$ line emission data of the J=1-0 transition from the FCRAO survey of the Taurus molecular cloud (Goldsmith et al. 2008) for another 9 of them.
The maps were centered on the coordinates given by Troland \& Crutcher (2008) and Falgarone et al. (2008) as referenced by CWHFT10. The angular sizes of the maps were 3 to 4 times the typical sizes of cores. The method we used to calculate the aspect ratios is based on the first and second moment of the flux density and is described in detail by Tassis et al. (2009). In certain cases, cores, as projected on the plane of the sky, are closely spaced; for this reason, flux from a nearby core can severely affect the aspect ratio of the object under examination. To avoid this, we performed a visual inspection to select a region that contains only one core. Finally, in order to remove the background which can also affect the analysis, we set a threshold and calculated the aspect ratio only from the pixels whose intensity was greater than the mean intensity of the final region (e.g., see Fig.~\ref{NGC2024} and inset). The resulting maps of the cores, along with their principal axes and their aspect ratios, are presented in Figures~\ref{multipanel_maps_250mu} - \ref{multipanel_maps_CO}.

Even visual inspection of these maps reveals that these objects do not appear to be spherical. Indeed, only 4 of these objects have an aspect ratio consistent with a spherical geometry. The mean value of the aspect ratios as computed here is 0.63, suggesting that in our sample of cores the preferred geometry is a flattened, oblate one  (Fig.~\ref{aspect_ratio_dist}). We therefore conclude that the shapes of the cores in the CWHFT10 sample are not consistent with the spherical geometry implied by the $B\propto \rho^{2/3}$ relation. 

In our theoretical analysis of shapes, we considered only the case of a pure dataset following a specific geometry and evolutionary path. It is also conceivable that a mixture of different object geometries, each with its own $B$ -- $\rho$ relation, could yield a value of $\kappa$ different from the values characterizing the individual objects, including possibly $\kappa = 2/3$. In order for such a scenario to be realized, objects with both smaller values of $\kappa$ (i.e., 1/2 or 0) and greater values of $\kappa$ (i.e., 1) need to be present in the sample. As discussed in \S \ref{geosection}, values of $\kappa$ greater than $2/3$ can be produced by one-dimensional contraction of a  disklike (oblate) cloud with its magnetic field parallel to the plane of the disk, or the lateral contraction of a long cylindrical cloud with its magnetic field along the cylinder (see $\S~2$). Both of these cases were found to be unlikely in a combined study of cloud shapes and magnetic fields using data for 32 clouds surveyed by the Hertz polarimeter (Tassis et al. 2009, see their Figs. 2b and 2e). The latter possibility is also contradicted by our study of the CWHFT10 core shapes alone, as it would require a different aspect ratio distribution than the one shown in Figure~\ref{aspect_ratio_dist}.  The CWHFT10 aspect ratio distribution is similar in shape to the one of the sample studied by Tassis (2007), and which was shown to be strongly peaked around intrinsically oblate cores. Even a uniform shape distribution (equal number of oblate and prolate objects) would require a much greater fraction of aspect ratios between $0$ and $0.4$. It is therefore unlikely, given what we know regarding core shapes and the orientation of their magnetic fields, that the CWHFT10 sample contains enough cores that evolve either as $B\propto\rho^{2/3}$ or as $B\propto\rho$ to effectively ``pull'' the observed log$B$ -- log$\rho$ relation away from a slope of $0.5$. 

\section{Tracing the source of the discrepancy between nonspherical shapes and CWHFT10's $\kappa = 2/3$}\label{tracing}

We have found in \S \ref{shapes} %and \S \ref{kappas}, 
that the preferred model in CWHFT10 would require spherical shapes, which are very rare among their observed objects.
In this section, we focus on tracing the cause of the disagreement between the results of the CWHFT10 analysis and the results of our own tests. 

\subsection{A joint treatment of H\small{I} and molecular clouds?}\label{joint}

One likely culprit is the set of assumptions regarding the {\em family of models} CWHFT10 used to describe their data. The conclusion of CWHFT10 about the $B - \rho$ relation is obtained through a careful Bayesian analysis. Despite the important conceptual and methodological advantages of Bayesian statistics in treating diverse datasets, Bayesian analyses are necessarily model-dependent: they select the best parameter values in a specific family of models to treat the data at hand; but they do not convey information on how good a fit to the data the adopted family of models is as a whole. For example, a Bayesian analysis can always find a set of parameters of a normal distribution that best describe a specific dataset, even if the data are not distributed normally. However, that ``best description'' is actually a very poor representation of reality.  

A central assumption in the CWHFT10 family of models that could be affecting their findings on the value of the exponent $\kappa$ is that concerning low densities (H\small{I} data), namely, that the average magnetic field remains constant with density. This assumption is problematic for two reasons.  First, observationally, it is not consistent with our {\it{empirical}} understanding of the behavior of magnetic fields with density for H\small{I} clouds. Several values of $\kappa$ have been proposed, but the value $\kappa = 0$ has not been one of them (e.g., see review by Vallee 1997). This has also been reported by Marchwinski et al. (2012), who explicitly contrasted their results to the model adopted by CWHFT10 at low densities. Second, if one forces data on the  $B$ -- $\rho$ plane on a horizontal line at low densities and demands continuity of the $B$ -- $\rho$ relation between low- and high-density datapoints, one necessarily sets a pivot point for the high-density part of the relation. If the value of the magnetic field at the ``transition density'' $n_0$ (the density which separates the low-density and high-density branches of the model) is too low, the value of $\kappa$ for the high-density part of the $B$ -- $\rho$ relation will be forced to acquire greater values in order to accommodate the magnetic fields at the highest densities.  

We devise a ``goodness of fit'' test to quantify whether indeed the assumption that the average magnetic field in H\small{I} clouds is independent of density is consistent with the data, and whether this assumption can affect the global model fit. 

We start with the set of observed number densities for the objects used in the CWHFT10 analysis. For each number density $n_{\rm i}$, we randomly select a magnetic field according to the best generalized model by CWHFT10. First we calculate the $B_{\rm max}$ value appropriate for $n_{\rm i}$ through, 
\begin{equation}
B_{\rm max}(n)=
\begin{cases}
B_0,\, \, \, \hfill n<n_0 \\
B_0 (\frac{n}{n_0})^\alpha, \, \, \, \hfill n>n_0
\end{cases}
\end{equation}
where $B_0$ = 10 $\mu$G, $n_0$ = 300 cm$^{-3}$, $\alpha$ = 0.65. We then select a total magnetic field $B_{\rm i}$ of the cloud from a uniform distribution with boundaries between $f\cdot B_{\rm max}$ and $B_{\rm max}$, as prescribed by CWHFT10.  To account for various orientations we multiplied each $B_{\rm i}$ by cos($\phi$) randomly drawn from a uniform distribution between $-1$ and $1$, to obtain the line-of-sight magnetic field $B_{\rm LOS, i}$. We then ``observed'' this $B_{\rm LOS,i}$ by drawing a random value from a normal distribution with mean equal to $B_{\rm LOS,i}$ and standard deviation equal to the actual observational uncertainty $\sigma_B$ recorded by CWHFT10 for the cloud with density $n_{\rm i}$. 

In this way, we produce ``mock'' magnetic field observations {\em at the same densities} as the CWHFT10 dataset that are consistent with the CWHFT10 model for the magnetic field. We then compare the mock data with the actual magnetic field measurements through a Kolmogorov-Smirnov test. We perform the test for (i) the entire dataset, (ii) just H\small{I} observations, and (iii) just molecular observations. When we treat the entire dataset (H\small{I} and molecular observations), the K-S $p$-value for the $\alpha = 0.65$ case was $19.7\%$ - the model is acceptable. However, when we treat the H\small{I} and molecular data separately, the picture changes: for the molecular observations alone (where $B$ {\em is} dependent on $\rho$ according to the CWHFT10 model), the $p$-value for $\alpha=0.65$ is only  $5.2\%$ - only marginally consistent with the data. In the case of H\small{I} data, the K-S test $p$-value is only $0.35\%$: $B\propto \rho^0$ is {\em not} a good description of the low-density data. 

\begin{figure*}
\includegraphics[width=2.0\columnwidth, clip]{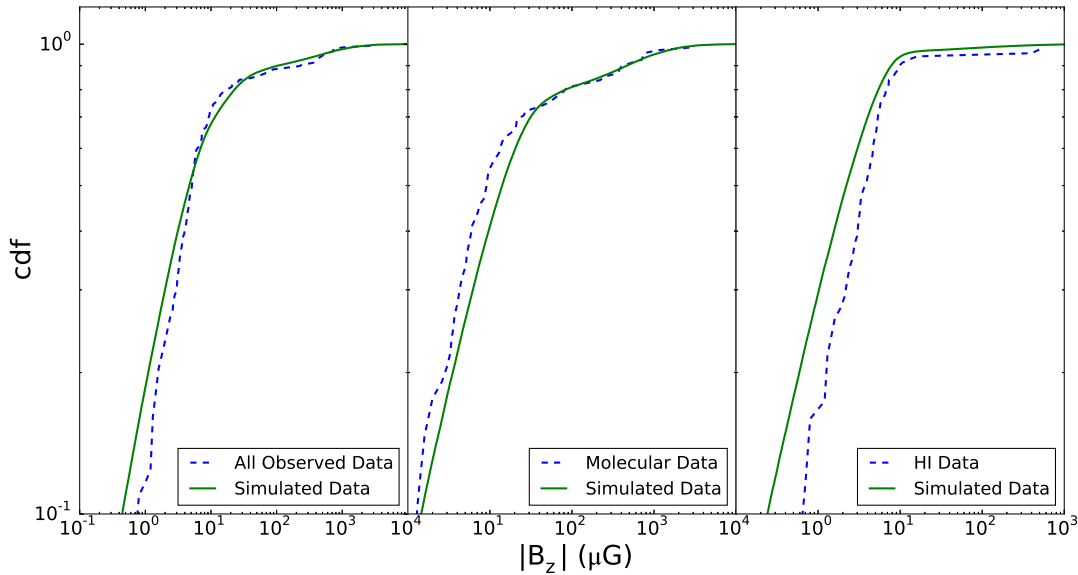}
\caption{Cumulative distributions of the actual observations (dashed blue lines) and mock-observed values drawn from the CWHTF10 ``best-fit'' model (green solid lines) of each of the following cases: entire dataset (left), just molecular observations (middle), and just HI observations (right).}
\label{CDFS}
\end{figure*}

Figure \ref{CDFS} shows the cumulative distribution functions (CDFs) of the measured line-of-sight value of the magnetic field, for the different datasets, and demonstrates what happens when we combine H\small{I} and molecular data: both H\small{I} and molecular datasets are in poor agreement with the model, but {\em in opposite directions}: one produces too few low values of $B$ (data CDF starts out below the model), and the other produces too many (data CDF starts out above the model). Adding the two datasets together moves the data CDF closer to the model. 

We have therefore shown quantitatively in this section that treating H\small{I} and molecular clouds jointly can affect the acceptability of a value of $\kappa$ for the molecular branch of the model. For this reason, for the remainder of this work, we treat the molecular data by themselves.

\subsection{An uncertainty of only a factor of 2 in volume densities?}\label{fac2}

\begin{figure*}
\center
\includegraphics[width=2.0\columnwidth]{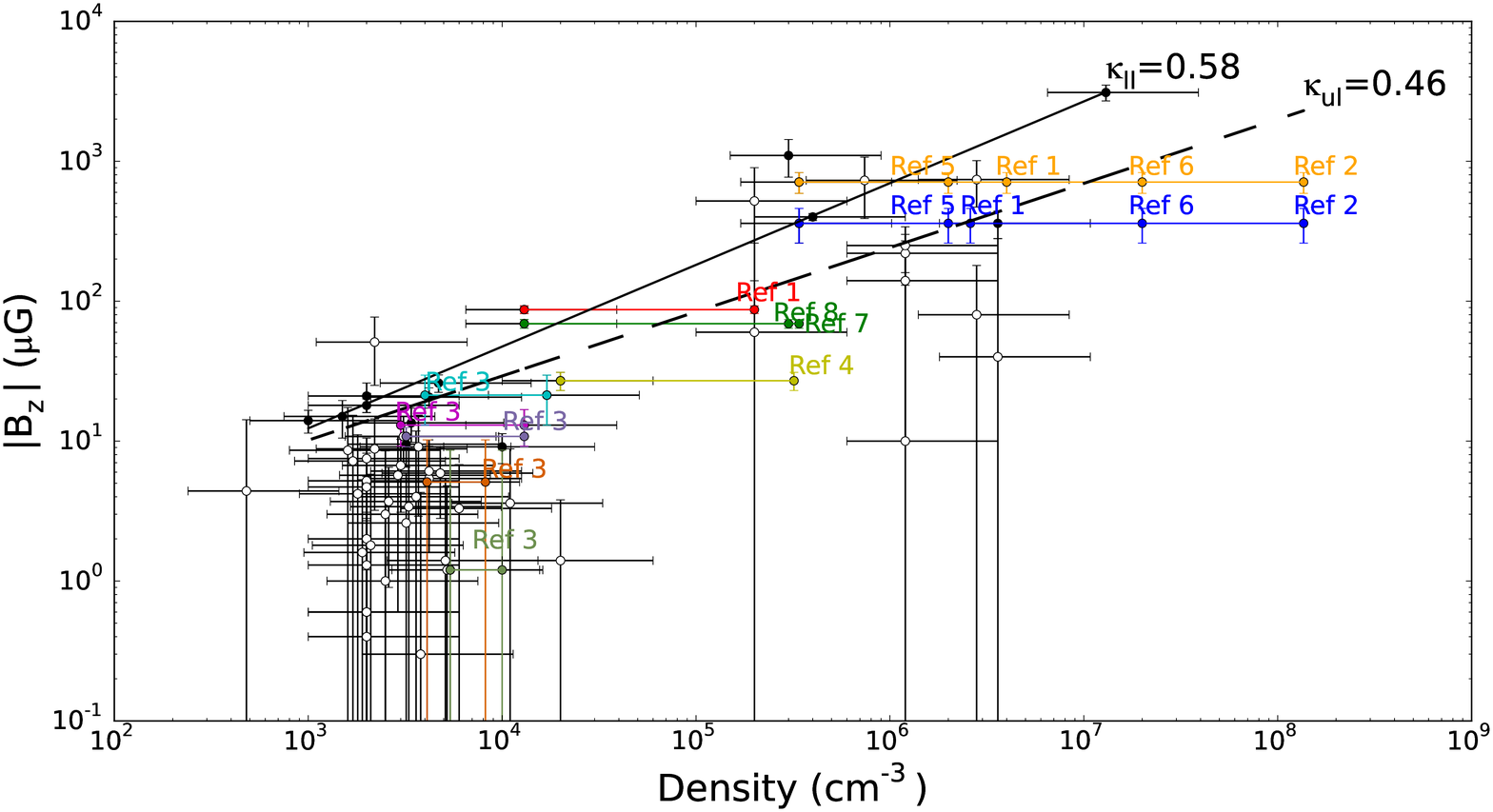}
\label{Brhoplot}
\caption{Two least-squares fits to detections only, when upper and lower limits of volume density measurements (curves labeled by $\kappa_{\rm{ul}}$ and $\kappa_{\rm{ll}}$, respectively) are adopted from the literature. Even when Zeeman measurements are accurate enough, volume density uncertainties can produce ambiguous results. Objects for which multiple measurements are available, and references, are given in Table \ref{denstable}. \label{uncertfig}}
\end{figure*}

\begin{table}
\centering
\caption{\label{denstable} Density measurements of various objects in CWHFT10. Values from: Crutcher et al. 1999 (1); Mookerjea et al 2012 (2); Troland \& Crutcher 2008 (3); Hirota et al. 1997 (4); Motte et al. 2007 (5); Girart et al 2013 (6); Sarma et al 2013 (7); Saito et al. 2007 (8).}
\begin{tabular} {c c c c}
\hline\hline \\
Core &CWHFT10 value & Other Measurements&Ref.\\
& $(\rm{cm}^{-3})$& $(\rm{cm}^{-3})$& \\
\hline \\
NGC 2024&$1.3\times10^{4}$&$2\times10^5$&(1)\\
\\
S88B&$1.3\times10^{4}$&$3\times10^{5}$&(8)\\
\\
&&$3.4\times10^{5}$&(7)\\
\\
DR21OH1&$3.4\times10^{5}$&$2\times10^{6}$&(1)\\
\\
&&$1.36\times10^{8}$&(2)\\
\\
&&$2\times10^{6}$& (5)\\
\\
&&$2\times10^{7}$&(6)\\
\\
DR21OH2&$3.4\times10^{5}$&$4\times10^{6}$&(1)\\
\\
&&$1.36\times10^{8}$&(2)\\
\\
&&$2\times10^{6}$&(5)\\
\\
&&$2\times10^{7}$&(6)\\
\\
L1457S&$1.3\times10^{4}$&$3\times10^{3}$&(3)\\
\\
L1457Sn&$1.7\times10^{4}$&$4\times10^{3}$&(3)\\
\\
L1534&$5.4\times10^{3}$&$1\times10^{4}$&(3)\\
\\
L1544&$3.2\times10^{3}$&$1.3\times10^{4}$&(3)\\
\\
L1595(6)&$4.1\times10^{3}$&$8.2\times10^{3}$&(3)\\
\\
B1&$2\times10^{4}$&$3.2\times10^{5}$&(4)\\
\hline\hline
\end{tabular}
\end{table}

An additional potentially problematic issue in the CWHFT10 analysis is the estimate of uncertainties in the density of each cloud/core.  In CWHFT10 the issue is stated as follows: ``Based on our experience in making such estimates, we choose to trust the inferred value within a factor of two, although the actual degree of uncertainty is not precisely known'' (Cructher et al. 2010). However, the uncertainty in the data is an important factor in any statistical analysis. Underestimated uncertainties can make a dataset appear much more constraining than it really is. For this reason, we try to assess the actual degree of uncertainty for as many objects as possible by conducting a literature search of density estimates for each object and examining the spread between different estimates. 

In approximately half of the total number of molecular datapoints given in Crutcher et al. (2010), volume densities are taken from surveys different from those from which the magnetic field data are taken. Even if the centre coordinates and angular resolution between the Zeeman survey and the one probing other physical parameters are the same, volume densities are usually derived from CO and CS measurements (see appendices in Crutcher 1999, and Falgarone et al. 2008). 
This introduces a bias since these molecules provide information for regions different from those where Zeeman emission occurs. Additional chemical assumptions, as for example the $N$(H$_2$)/$N$(CO) ratio and the depletion of the species onto dust grains, can further complicate the picture (see Tassis et al. 2012; 2014). 

An example of the spread of volume-density estimates using different tracers is DR21OH: Vall\'{e}e $\&$ Fiege (2006) used $^{12}$CO observations to determine the volume density of the core, which they reported to be $n(\rm {H_2}) = 3.4 \times 10^5 ~ {\rm cm^{-3}}$. Mangum et al (1991)  used C$^{18}$O observations to derive $n(\rm {H_2}) \ge 6 \times 10^6 ~{\rm cm^{-3}}$ for the same object.  The difference is over an order of magnitude, much greater that the factor of 2 used by CWHFT10. Further uncertainties are introduced through the morphology of the cores. In absence of a better practice, column densities are converted to volume densities by assuming spherical geometry. However, no real information exists for the line-of-sight size of the core. It is clear from the above that the uncertainties in the volume densities can be much greater than the factor of two adopted by CWHFT10. 

Studying the cited literature, we find that volume-density values reported in CWHFT10 are not always consistent with the references to which they are attributed, presumably as a result of updated data analysis (cores NGC2024, L1457S, L1457Sn, L1495(6), L1534, L1544). In any case, the changes in the adopted volume densities are indicative of the uncertainties introduced by various choices in the data analysis.

Even if discrepancies greater than a factor of two are only present in a small number of cases, since the total number of points is small, the resulting effect on the statistical analysis of the dataset can be significant. We find the greatest variations for the more evolved cores; measurements of volume densities in other literature sources are generally greater than the ones reported in CWHFT10. Table \ref{denstable} summarizes different volume density measurements for the fraction of CWHFT10 objects for which a literature search yielded additional density estimates. Figure (\ref{uncertfig}) demonstrates the effect of density uncertainties on the slope of the $B$ -- $\rho$ relation in a simple way, focusing on datapoints with significant ($>3\sigma$) measurements of the magnetic-field strength rather than upper limits. Objects for which multiple measurements of the volume density are available in the literature are shown in color. If the lowest available measurements of the volume density are used, the value of $\kappa$ (apparent slope derived through log-log regression on the $B$ -- $\rho$ plane using detection-only data) is $\kappa_{\rm ll}$ = 0.58 (solid line). These points are consistent with the volume-density values adopted by CWHFT10. If the highest available measurements of the volume density are used, then $\kappa_{\rm ul}$ = 0.46 (dashed line). 

\begin{figure}
\begin{center}
\includegraphics[width=1.0\columnwidth, clip]{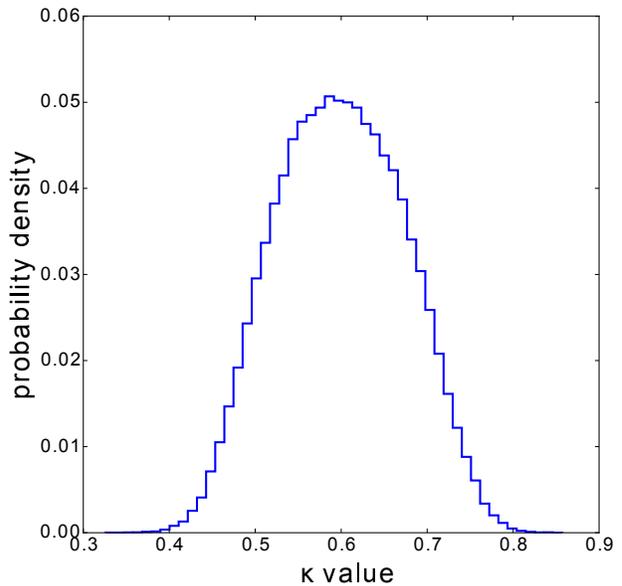}
\caption{Distribution of apparent slope $\kappa$ derived through log-log regression on the $B$ -- $\rho$ plane of detection-only data for the optimal model of CWHFT10, using updated volume-density uncertainties. 
\label{k_dist}}
\end{center}
\end{figure}

We can also quantify the statistical effect of the updated uncertainties, by performing a Monte-Carlo simulation of the apparent slope $\kappa$ derived through a naive statistical analysis of mock observations drawn from the CWHFT10 model with  updated uncertainties in the volume density where these are available (and a factor of 2 when no additional information is available). Our analysis is similar to that of Crutcher (1999): a single power law is fitted through regression to the detections-only data (as defined in Crutcher 1999: measured line-of-sight $|B|$ in units of its observational uncertainty greater or equal than 2.5). We repeat the ``experiment'' $10^5$ times. The  distribution of $\kappa$ values obtained in this manner is shown in Figure~\ref{k_dist}. A Gaussian fit to the distribution shown yields a mean of 0.59 and a spread of 0.07: the detections-only slope is nonconstraining, and can be anywhere from $\sim 0.4$ to $\sim 0.8$ within 2$\sigma$ (assuming the CWHFT10 model is correct). 

We conclude that there are several cases for which the true uncertainty in the volume density is much greater than the factor of 2 adopted by CWHFT10, and that the effect of changing the adopted volume-density measurements within the range of estimates found in the literature can have a very significant effect on the $B$ -- $\rho$ relation of at least detections-only data. 

In order to quantify the effect of the large volume-density uncertainties in the treatment of the entire datasets, we repeat the goodness-of-fit test of \S \ref{joint} for molecular objects only, properly accounting for uncertainties in density: for objects for which multiple volume-density estimates exist in the literature, we choose a density uniformly distributed in the available range. For other objects, we assign a density within a factor of two of the CWHFT10 value. We calculate the cumulative distribution of simulated (mock-observed) magnetic-field values and we check its consistency with the cumulative distribution of actual magnetic-field strength observations through a K-S test. The hypothesis that the two distributions are the same is rejected at the 0.3\% level. 

\begin{figure}
\center
\includegraphics[width=1.0\columnwidth, clip]{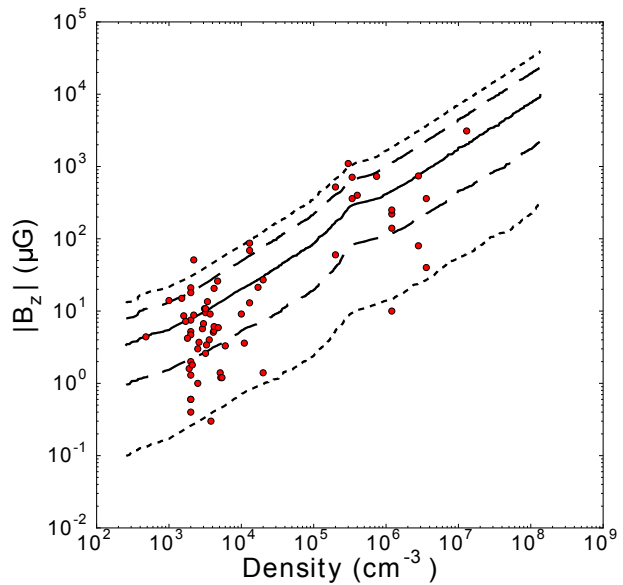}
\caption{Median (solid line), $1\sigma$ (dashed lines) and $2\sigma$ (dotted lines) limits of the simulated $|B_z|$ at each density according to the CWHTF10 model, overplotted with molecular data from CWHTF10.
\label{clines}}
\end{figure}

The reason for this discrepancy can be seen in Figure \ref{clines}, which overplots data and model predictions on the $B$ -- $\rho$ plane. Using the same algorithm described above, we calculate, at each density bin, the median simulated $|B_z|$ (solid line)  as well as the $|B_z|$ limits that contain $1\sigma$ ($68\%$, dashed lines) and $2\sigma$ ($95\%$, dotted lines) of our simulated observations. On the same plot, we overplot with red dots the measurements that come from molecular tracers in CWHFT10. The cause of the strain registered by the K-S test between the CWHFT10 model and the data can be clearly identified in this plot. While reasonable fractions of points are within the expected $1\sigma$ and $2\sigma$ limits at various densities, they are not symmetric about the expected median: {\it a larger fraction of points lies systematically below the median than above}.

\subsection{A flat distribution of magnetic field strengths?}\label{lognormal}

A final potentially problematic assumption in CWHFT10 is that of the shape of the probability distribution of magnetic-field values. CWHFT10 adopt a uniform distribution of $B$-values, between some minimum and some maximum value. There are two issues arising from this choice. The first one is statistical. 
The uniform distribution is a convenient tool and significantly simplifies statistical analyses when it can be used without loss of generality or in absence of further information. However, it is always the most restrictive option among all frequently used bi-parametric distributions of the same variance $\sigma^2$. The reason is that, for a uniform distribution of finite width, the probability to have intrinsic (i.e., free of observational error) values of the quantity of interest at a distance greater than $\sqrt{3}\sigma$ from the mean is exactly zero. This means that this type of distribution cannot adequately describe peaked distributions with tails. If a collection of datapoints that shows a preferred value (a peak) as well as outliers (tails) is forced onto a uniform distribution of variable width, the best fit that will be obtained by likelihood analysis is a uniform distribution that completely misses the peak and stretches out all the way to the farthest outlier. In terms of the $B$ -- $\rho$ scaling problem: in order to accommodate, say, a single abnormally high value of the $B$--field at very high densities, the uniform $B$-distribution has to stretch its maximum to very high values, and this could again result in a steeper scaling than otherwise warranted by the data. This effect of the uniform-distribution choice is somewhat moderated by the treatment of observational uncertainties (making the probability of observation of an outlier finite due to observational error), but it is not clear that this is enough to eliminate the bias that could potentially be introduced. 

The second problem with the choice of a uniform distribution of $B$-values is conceptual. CWHFT10 assign a physical interpretation to the width of the best-fit uniform distribution they derive from their data: that, since the best-fit distribution extends uniformly from almost zero to a maximum value, this implies that there is no preferred value of the magnetic field in objects of a specific density, and that this is additional evidence for the dynamical insignificance of magnetic fields, compounded with their preferred $2/3$ slope of the $B$ -- $\rho$ scaling. However, if there {\em were} a preferred value of the magnetic field in the interstellar medium, but there were also outliers, the CWHFT10 uniform distribution would miss it and, consequently, would lead to a misleading physical interpretation of the available data. 

The first question we ask in addressing these issues is whether a different family of distributions which {\em is} capable of describing peaked distributions with tails may be an equally good or better fit to the magnetic-field data. Such a family of distributions is the lognormal, 
\begin{equation}
p(B) = \frac{1}{B\sqrt{2\pi}\sigma_0}\exp\left[-\frac{(\ln B - \ln B_{\rm m,a})^2}{2\sigma_0^2}\right] ,\ 
\end{equation}
which represents a Gaussian distribution of $\ln B$ centered at $\ln B_{\rm m,a}$ and with spread $\sigma_0$.
We will identify optimal parameters for a lognormal (instead of a uniform) distribution of intrinsic magnetic-field strengths, $B$, which is consistent with the observed $|B_z|$ values in a narrow density range (from $2\times 10^3$ to $4\times 10^3 {\rm \, cm^{-3}}$). We have chosen that particular density range because it is well populated (27 observed objects) so that it provides enough statistics to give a good sense of the underlying distribution, and narrow enough so that we can ignore any evolution with density between points.

In order to determine parameters of this distribution, we scan the two-dimensional parameter space of ($B_{\rm m,a}$, $\sigma_0$), and for each pair we simulate a distribution of observed $|B_z|$ with random observation directions (uniform $\cos \theta$) and including Gaussian observational errors identical to the ones quoted in CWHTF10. The resulting distribution is then compared to the distribution of observed $|B_z|$. Since the purpose of this work is not to repeat the sophisticated statistical analysis of CWHTF10 but simply to assess the effect of relaxing the potentially problematic assumptions, we have not formally obtained a best-fit distribution; instead, we selected the set of parameters that minimizes the K-S statistic (yielding, in our case, a K-S $p$-value of 80\% of the two datasets to be drawn from the same distribution). This means that a proper fit (for example, through a maximum-likelihood analysis) may yield a slightly different set of optimal parameters, which, if anything, will be an {\em even better} fit to the data. The optimal parameters we have identified for the lognormal in this narrow density range are $\ln (B_{\rm m,a}/{\rm \mu G}) =2.57$ and $\sigma_0 =0.3$, respectively. We assign this distribution to a number density of $n_{\rm a}=2.61\times 10^3 {\rm \, cm^{-3}}$, which is the average of the CWHTF10 quoted densities for objects in the density range we have considered.  In the analysis that follows we assume that, for all densities above $300 {\rm \, cm^{-3}}$, the distribution $p(B)$ remains lognormal with the same $\sigma_0$ at all densities, while $B_{\rm m}$ scales as 
\begin{equation}
B_{\rm m}(n) = B_{\rm m,a}(n_{\rm a})\left(\frac{n}{n_{\rm a}}\right)^\alpha\,.
\end{equation}

\begin{figure}
\includegraphics[width=1.0\columnwidth, clip]{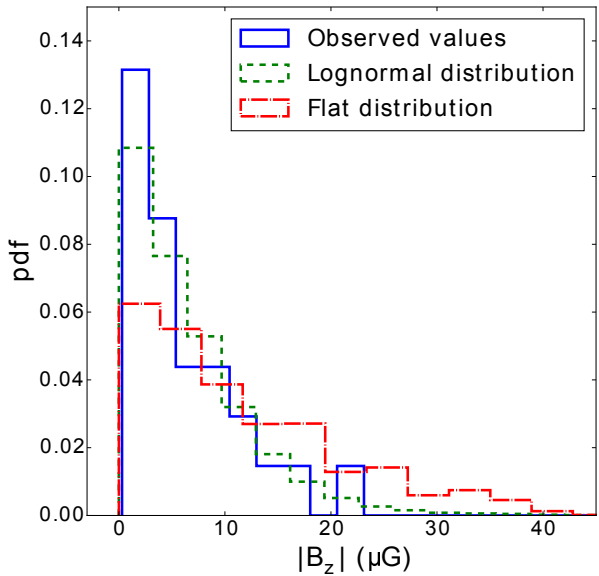}
\caption{Distribution of line-of-sight magnetic field values $B_z$ in the density range $2-4\times 10^3 {\rm \, cm^{-3}}$. Blue histogram (solid line): all observed objects (molecular data only). Green histogram (dashed line): optimal lognormal distribution. Red histogram (dashed-dotted line): CWHTF10 model (uniform distribution).
\label{Bzpdfs}}
\end{figure}

Figure \ref{Bzpdfs} shows how the observed $|B_z|$ values in the density range we have considered compare to those expected from the optimal lognormal $B$ distribution and the CWHTF10 uniform $B$ distribution (assuming in both cases uniform viewing angles $\cos \theta$ and observational uncertainties as quoted in CWHTF10). It is clear that the lognormal is a better match for the qualitative behavior of the observed data. 

\begin{figure}
\includegraphics[width=1.0\columnwidth, clip]{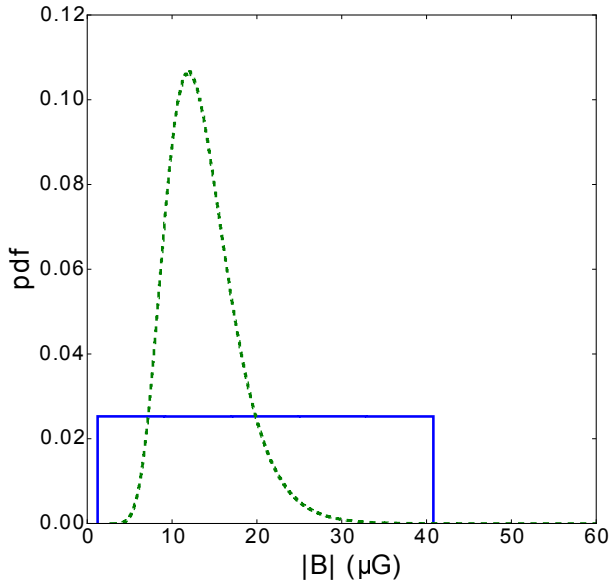}
\caption{Distribution of intrinsic magnetic-field strength $B$ at $n = 2.61 \times 10^3 {\rm \, cm^{-3}}$. Green (dashed) line: optimal lognormal; Blue (solid) line: CWHTF10 model (uniform distribution).
\label{Bpdfs}}
\end{figure}

Figure \ref{Bpdfs} shows the corresponding probability density functions for the intrinsic magnetic-field strength $B$ (instead of the observed line-of-sight strength of the magnetic field, $|B_z|$) at the number density $n_{\rm a}$. The plot demonstrates how the CWHFT10 conclusion, that there appears to be no preferred value of the magnetic field at a specific density in the interstellar medium, because a wide uniform distribution is preferred over a much narrower one, is a direct result of the constraining nature of a uniform distribution: the optimal lognormal, which we have shown is a better qualitative description of the observed data, {\em does} show a significant peak, corresponding to a preferred value at a given density; however, significant tails exist. In contrast, the optimal uniform distribution misses this peak, and extends to high and low values in order to accommodate the tail values, giving the false impression that the spread in intrinsic $B$ needed to explain the observed $B_z$ is much greater than it is in reality. 

\section{Reconciling the $B$ -- $\rho$ relation with core shapes.}\label{reconciling}

Having identified three potentially problematic assumptions in the CWHFT10 analysis, we now re-evaluate the information their $(B_z,n)$ datapoints convey regarding the slope of the $B$ -- $\rho$ relation and we reconcile the latter with the lack of evidence for a preference for spherical geometry in the CWHFT10 cores. We do so by: (a) treating molecular (high-density) data on their own, since this is where the debated scaling ($B\propto \rho^{2/3}$ vs $B\propto \rho^{1/2}$) arises; (b) using updated uncertainties in the volume density where these are available; and (c) using a lognormal distribution of intrinsic magnetic-field strengths. 

\begin{figure}
\includegraphics[width=1.0\columnwidth, clip]{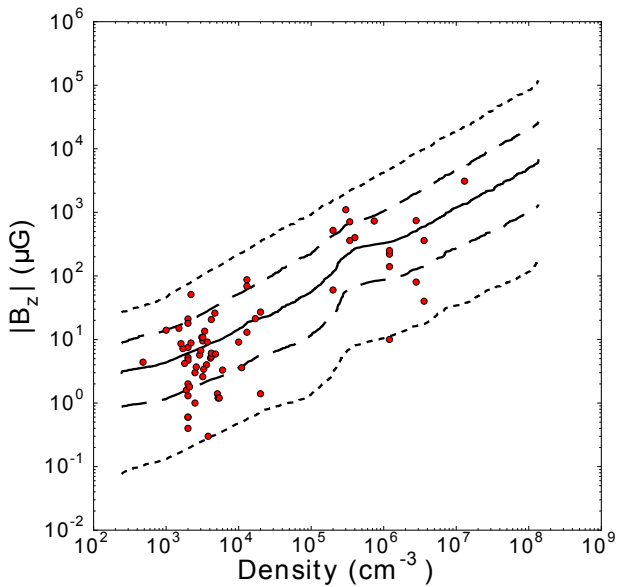}
\caption{Median (solid line), $1\sigma$ (long-dashed lines) and $2\sigma$ (short-dashed lines) limits of the simulated $|B_z|$ at each density obtained from the optimal lognormal $p(B)$ and $\alpha = 0.65$, overplotted with molecular data from CWHTF10.
\label{lines_65}}
\end{figure}

\begin{figure}
\includegraphics[width=1.0\columnwidth, clip]{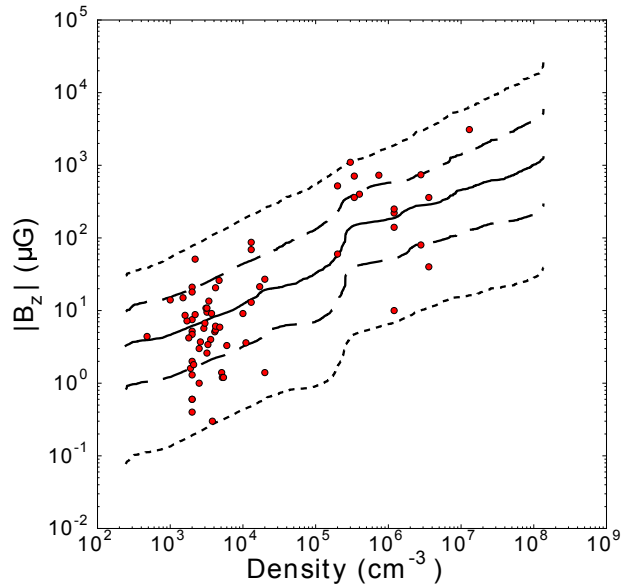}
\caption{Median (solid line), $1\sigma$ (long-dashed lines) and $2\sigma$ (short-dashed lines) limits of the simulated $|B_z|$ at each density obtained from the optimal lognormal $p(B)$ and $\alpha = 0.5$, overplotted with molecular data from CWHTF10.
\label{lines_5}}
\end{figure}

We repeat our goodness-of-fit tests of \S \ref{joint} using the optimal lognormal distribution described in \S \ref{lognormal}, with its mean scaled with $n$ with a slope $\alpha$ equal to either $0.65$ (the CWHFT10 preferred value) or $0.5$ (the historically preferred value due to theoretical expectations from magnetically-controlled gravitational contraction and due to the empirical results from fitting a power law to detection-only data). 
Figures \ref{lines_65} and \ref{lines_5} show the $|B_z|$ -- $n$ plane with lines corresponding to the median (solid), $1\sigma$ (long dashes), and $2\sigma$ (short dashes) expected limits of mock observations drawn from the lognormal model, with slope $\alpha$ equal to $0.65$ and $0.5$, respectively. A K-S test between the observed and simulated distributions of $|B_z|$ values in the entire density range of molecular datapoints, as in \S \ref{fac2}, returns a $p$-value of 7.5\% for observed and simulated data to be drawn from the same distribution, for a scaling slope of $0.65$; it returns a $p$-value of 15.2\% for a scaling slope of $0.5$. 


The agreement between data and model has improved significantly even for a scaling slope of 0.65 (from $0.3\%$ to $7.5\%$) by switching from a uniform distribution of $B$-values to a lognormal. This is achieved without changing the number of modeling parameters; both uniform and lognormal are bi-parametric families of distributions. However, a slope of $\alpha=0.5$ is preferred over $\alpha=0.65$, although the two cannot be distinguished at a statistically significant level. 

\vspace{-3ex}
\section{Summary}\label{sum}

In this paper, we discussed ways to assess the relation between the magnetic-field strength and the gas density in the interstellar medium.

We reviewed the connection between the exponent $\kappa$ in the relation $B \propto \rho^{\kappa}$ and the geometry of a cloud. We showed that specific combinations of cloud geometry and magnetic-field orientation result in different $B$ -- $\rho$ relations, with $B \propto \rho^{2/3}$ being unique to the spherical-cloud (or core) geometry. 

In light of this result, we sought to verify whether the claim of $B\propto \rho^{2/3}$ (at high densities) through the statistical analysis of a large sample of density and magnetic-field measurements by CWHFT10 is consistent with the geometry of objects in their sample. To this end, we have studied emission maps in all 27 objects in the CWHFT10 sample for which such data were available. We only found aspect ratios consistent with a spherical geometry in 4 of them; the distribution of aspect ratios does not show any preference toward unity, which would be the signature of spherical shapes. We thus concluded that there is no evidence of preferentially spherical objects in the CWHFT10, a result inconsistent with a $B\propto \rho^{2/3}$ relation. 

We then investigated the possibility that this disagreement could be caused by simplifying assumptions in the CWHFT10 statistical analysis. 

We first tested the effect of using a joint model for low- and high-density data and forcing the low-density data to a constant magnetic-field strength, independent of density. We used Monte-Carlo simulations to produce mock observations from the best CWHFT10 model with identical observational uncertainties as the ones quoted in CWHFT10, and we used a Kolmogorov-Sminrov test to determine whether the mock observations were consistent with the data. We found that:
\begin{itemize}
\item When treating the two branches (H\small{I} and molecular) of the CWHFT10 model separately, the low-density $B\propto \rho^{0}$ branch is inconsistent with the data ($p$-value = 0.35\%), while the high-density branch is marginally consistent ($p$-value = 5.2\%). 
\item These two branches deviate from the data in opposite directions, and, as a result, when we treat the combined dataset, we find artificially improved consistency ($p$-value = 19.7\%). 
\end{itemize}

We therefore conclude that the finding of CWHFT10, that a model with $B\propto\rho^0$ at low densities and $B\propto \rho^{2/3}$ at high densities is the preferred description of observations of densities and magnetic fields, is an artifact of their combined treatment of low- and high-density datapoints. 

We also checked whether the uncertainty in volume densities of a factor of 2 adopted by CWHFT10 is a good estimate, by tracing literature sources and comparing different density estimates for the same objects when these were available. We found that for several objects the actual uncertainties, as reflected in the spread of estimates in the literature, are much greater. When repeating our consistency analysis between the high-density branch of the CWHFT10 model and the data using updated volume-density uncertainties, the agreement between model and data worsens, with the $p$-value dropping to 0.3\%. The reason for this is that the additional density values available in the literature (especially at the highest-density objects) tend to be greater, instead of being symmetrically distributed about the value adopted in CWHFT10: more recent estimates have generally produced upward corrections in volume densities. 

We investigated whether a lognormal distribution for $p(B)$ would yield better agreement with the data than the CWHFT10 uniform distribution. We found that it does. 

Finally, we relaxed the three problematic assumptions, by (a) treating molecular observations on their own, (b) using updated volume-density uncertainties, and (c) using the optimal longormal to model the distribution of magnetic-field strengths. We repeated our goodness-of-fit K-S tests for the lognormal models, scaled with density with slopes of $0.65$ (the optimal CWHFT10 slope) and $0.5$. We found that the K-S test accepts both models. However there is a preference for $B\propto\rho^{1/2}$ (by a factor of 2 in the $p$-value of the K-S test), which is also preferred by the independent analysis of cloud shapes. This result is in agreement with predictions of the ambipolar-diffusion theory of star formation.

\vspace{-4ex}
\section*{Acknowledgements}

KT acknowledges support by FP7 through Marie Curie Career Integration Grant PCIG- GA-2011-293531 “SFOnset”.
AT, GVP, and KT acknowledge partial support from the EU FP7 Grant PIRSES-GA-2012-31578 “EuroCal”.
GVP, KT, and TM  acknowledge support by the “RoboPol” project, which is implemented under the
“ARISTEIA” Action of the “OPERATIONAL PROGRAMME EDUCATION AND LIFELONG LEARNING” and is co-funded by the European Social Fund (ESF) and Greek National Resources. TM is grateful for the hospitality of the Department of Physics of the University of Crete during part of this work and writing of this paper. AT, GP, and KT thank the Astronomy Department of Caltech, for its hospitality during various stages of this work. We have made ample use of the public \textit{Herschel} Science Archive data products. We thank P. Goldsmith for providing the CO data of the Taurus Molecular Cloud and for comments on the manuscript. We are grateful to T. Hartquist for careful reading of the paper and insightful comments.

\end{document}